\documentclass[aps,prl,groupedaddress]{revtex4-1}
\usepackage{graphicx}
\usepackage{amsmath} 

\begin{document}


\title{Non-standard FDTD implementation of the Schr\"odinger equation}%
\author{J. M. N\'apoles-Duarte}%
\email{jnapoles@uach.mx}
\author{M. A. Chavez-Rojo}%
\affiliation{Facultad de Ciencias Qu\'imicas, Universidad Aut\'onoma de Chihuahua, Nuevo campus universitario, circuito universitario,  Apartado Postal
669, Chihuahua, Chihuahua, 31125 M\'exico.}




\date{\today}

\begin{abstract}
In this work, we apply the Cole's non-standard form of the FDTD to solve the time dependent Schr\"odinger equation. We deduce the equations for the non-standard FDTD considering an electronic wave function in the presence of potentials which can be higher or lower in comparison with the energy of the electron. The non-standard term is found to be almost the same, except for a sine function which is transformed to a hyperbolic sine function, as the argument is imaginary when the potential has higher energy than the electron.
Perfectly Matched Layers using this methodology are also presented.
\end{abstract}


\maketitle

\section{Introduction}
The Finite-Difference Time-Domain method (FDTD) was originally developed by Yee in 1966\cite{Yee} as a tool to solve complex problems in electromagnetics, and many advances has been achieved since then. One of the first attempts to apply the same method to solve the Schr\"odinger equation was proposed by Sullivan in his book.\cite{SullivanBook} Since then, several works have been published with substantial improvements.\cite{Sullivan2003,Soriano2004,Sudiarta2007,Sullivan2012} Here we propose to use the Non-Standard formalism presented by Cole\cite{Cole2013} to the quantum mechanical version of the FDTD.

\section{The FDTD for the Schr\"odinger equation}
Before starting with the non-standard form, its ought to recall the well stated FDTD approach to solve the one-dimensional time dependent Schr\"odinger equation:
\begin{equation}\label{eq:sch}
i\hbar\frac{\partial \psi (x,t)}{\partial t}=-\frac{\hbar^2}{2m}\frac{\partial^2\psi(x,t)}{\partial x^2} + V(x)\psi (x,t),
\end{equation}
where
\begin{equation}
\psi (x,t)=\psi_{real} (x,t)+j\psi_{imag} (x,t)
\end{equation}
with $j$ being the imaginary unit, and $\psi_{real}$ and $\psi_{imag}$ are of course the real and imaginary parts of the complex wavefuction. Visscher \cite{Visscher1991} proposed to split the Schr\"odinger equation by explicitly separating the real and imaginary parts of the wave function. Sullivan \cite{SullivanBook} used the same approach to write eq. \ref{eq:sch} into two coupled equations: 
\begin{subequations}
\begin{align}
\frac{\partial \psi_{real}(x,t)}{\partial t} &=-\frac{\hbar^2}{2m}\frac{\partial^2\psi_{imag}(x,t)}{\partial x^2}+\frac{1}{\hbar}V(x)\psi_{imag}(x,t),\label{eq:s1} \\
\frac{\partial \psi_{imag}(x,t)}{\partial t} &=\frac{\hbar^2}{2m}\frac{\partial^2\psi_{real}(x,t)}{\partial x^2}-\frac{1}{\hbar}V(x)\psi_{real}(x,t), \label{eq:s2}
\end{align}
\end{subequations}
this has the advantage of avoiding the use of complex numbers. 
The FDTD method uses the central difference approximation, in which the partial differential equations are discretized as follows. First, we have to approximate the space derivative by the central difference model:
\begin{equation}
\frac{\partial  \psi }{\partial x} \simeq \frac{d_x \psi^n(i)}{\Delta x},
\end{equation}

where we are using the notation $\psi^n(i)=\psi(i\cdot \Delta x, n\cdot \Delta t)$ commonly used in the literature,  $\Delta x$ and $\Delta t$ are the space and time steps, and $d_x\psi^n(i)=\psi^n(i+1/2)-\psi^n(i-1/2)$. A very similar equation can be easily obtained for the first order time derivatives.
Then, as we are dealing with laplacians, we also need to approximate the second order derivative:
\begin{equation}
\frac{\partial^2 \psi}{\partial x^2}\simeq \frac{d_x^2 \psi^n(i)}{\Delta x^2},
\end{equation}
with $d_x^2\psi^n(i)=\psi^n(i+1)-2\psi^n(i)+\psi^n(i-1)$. Finally, applying these approximations, the update finite difference equations obtained from eqs. \ref{eq:s1} and \ref{eq:s2} are respectively:
\begin{subequations}
\begin{align}
\psi_{real}^{n}(i+1/2) &=\psi_{real}^{n-1}(i+1/2)-\frac{\hbar \Delta t}{2 m (\Delta x)^2 }d_x^2\psi_{imag}^{n-1/2}(i) +\frac{\Delta t}{\hbar}V(i)\psi_{imag}^{n-1/2}(i),\\
\psi_{imag}^{n+1/2}(i) &=\psi_{imag}^{n-1/2}(i)+\frac{\hbar \Delta t}{2 m (\Delta x)^2}d_x^2\psi_{real}^{n}(i+1/2) -\frac{\Delta t}{\hbar}V(i+1/2)\psi_{real}^{n}(i+1/2).
\end{align}
\end{subequations}
As we can see, real and imaginary parts of the wave function are time and space shifted by half a step. This is the core of the FDTD method and when implemented in a computer code, it gives the time evolution of the wavefunction as the time evolves in a loop where the variable $n$ runs implicitly. Numerical errors are diminished with appropriate election of the space and time steps ($\Delta x$ and $\Delta t$), which are typically smaller than wavelengths in the simulation. Stair casing is a common source of artifacts that can be remediated by brute force taking $\Delta x$ and $\Delta t$ smaller until an acceptable result is obtainded, but at the cost of a larger computational effort.

Some stability issues need to be care before applying the computational algorithm, but here we take the case proposed by Sullivan,\cite{SullivanBook} {\em i.e} that the relationship between the spatial and temporal steps is
\begin{equation}\label{eq:dxdt}
 \frac{\Delta t}{\Delta x^2} \left( \frac{\hbar}{2m} \right)=\frac{1}{8}.
\end{equation}

The previous equations have been proved to model well some typical quantum mechanics problems, like quantum wells and  quantum dots. 
\section{NS-FDTD}
Cole \cite{Cole2013} developed a Non-standard variant for finite difference approaches, called non-standard FDTD (NS-FDTD). In this methodology, it was proposed the substitution of $\Delta x$ by some function $S(\Delta x)$. To obtain $S(\Delta x)$ we can assume monochromatic plane waves $\psi=\exp i(kx\pm\omega t)$ propagating in space. Thus
\begin{equation}
\frac{\partial}{\partial x}\psi= \frac{d_x \psi}{S(\Delta x)},
\end{equation}
using the discretized form of $\psi$, it is easy to find that
\begin{equation}
S(\Delta x)=\frac{2\psi^n(j)\sin(k\frac{\Delta x}{2})}{k\psi^n(j)},
\end{equation}
or 
\begin{equation}
S(\Delta x)=\frac{2}{k}\sin(\frac{k\Delta x}{2}).
\end{equation}

As we know from quantum mechanics, the momentum is related to the energy and frequency. For a more complex problem, we need to study
 the wave propagation as a function of the wave energy and the potential where it travels. 
\subsection{Case $E<V$}
 Let us first start with the case $E<V$.
 As it is known, with this condition, $k=i\sqrt{2m|E-V|}/\hbar$. For convenience we define $\gamma=\sqrt{2m|E-V|}/\hbar$, thus $k=i\gamma$. Then, we can find that 
 \begin{equation}
 S(\Delta x)=\frac{2}{\gamma}\sinh{\left( \frac{\gamma\Delta x}{2}\right)},
 \end{equation} 
where $\sinh$ is the hyperbolic sine function.
 On the other hand, doing a similar procedure, we found that
   \begin{equation}\label{eq:dt1}
 S(\Delta t)=\frac{2}{\left(\frac{V-E}{\hbar} \right)}\sin{\left(\left[\frac{V-E}{\hbar} \right] \frac{\Delta t}{2}\right)},
 \end{equation}
 or in an alternative form:
    \begin{equation}
 S(\Delta t)=\frac{2}{\gamma^2}\left(\frac{2m}{\hbar}\right)\sin{\left(\left[  \frac{\gamma \Delta x}{4}\right]^2 \right)}
 \end{equation}
 where we have used eq. \ref{eq:dxdt}.
 
 \subsection{Case $E>V$}
 For this case, we use $k=\sqrt{2m|E-V|}/\hbar$, and in a very similar fashion, we can find $S(\Delta x)$ and $S(\Delta t)$. First, we have
 \begin{equation}
 S(\Delta x)=\frac{2}{k}\sin{\left( \frac{k\Delta x}{2}\right)},
 \end{equation} 
 and
\begin{equation}\label{eq:dt2}
 S(\Delta t)=\frac{2}{\left(\frac{E-V}{\hbar} \right)}\sin{\left[\left(\frac{E-V}{\hbar} \right) \frac{\Delta t}{2}\right]}
\end{equation}
(Notice that eqs. \ref{eq:dt1} and \ref{eq:dt2} are completely equivalent.) In alternative form, we can write:
   \begin{equation}
 S(\Delta t)=\frac{2}{k^2}\left(\frac{2m}{\hbar}\right)\sin{\left(\left[  \frac{k \Delta x}{4}\right]^2 \right)}.
 \end{equation}
\subsection{Iterative Equations}
In the non-standard formulation of the FDTD, the so-called NS-FDTD, can be written as
\begin{equation}
\frac{\psi^n_{real} (j)-\psi^{n-1}_{real} (j)}{S(\Delta t)}=-\frac{\hbar}{2m}\frac{d_x^2\psi^{n-1/2}_{imag} (j)}{S(\Delta x)^2} + \frac{V(j)}{\hbar}\psi^{n-1/2}_{imag} (j)
\end{equation}

We define the non-standard term as:
\begin{equation}
  U_{NS}=\left\{
    \begin{array}{ll}
      \frac{\sin\left( \left[ \frac{\gamma \Delta x}{4} \right]^2 \right)}{\left[ \sinh\left( \frac{\gamma \Delta x}{2} \right)\right]^2}
&     E<V. \\
\\
      \frac{\sin\left( \left[ \frac{k \Delta x}{4} \right]^2 \right)}{\left[ \sin\left( \frac{k \Delta x}{2} \right) \right]^2} &   E>V.

    \end{array}
  \right.
\end{equation}
Notice that as $U_{NS}$ depends on $V$, then  $U_{NS}=U_{NS}(i)$.
The following iterative equations can be used to implement the NS-FDTD for both cases:
\begin{equation}\label{eq:ns1}
\psi^n_{real} (i+1/2)=\psi^{n-1}_{real} (i+1/2)-c_1(i)   d_x^2\psi^{n-1/2}_{imag} (i)  + c_2(i)V(i) \psi^{n-1/2}_{imag} (i)
\end{equation}
and,
\begin{equation}\label{eq:ns2}
\psi^{n+1/2}_{imag} (i)=\psi^{n-1/2}_{imag} (i)+c_1(i+1/2)   d_x^2\psi^{n}_{real} (i+1/2)  - c_2(i+1/2)V(i+1/2) \psi^{n}_{real} (i+1/2)
\end{equation}
where we define
\begin{equation}\label{eq:c1}
c_1(i)=\frac{U_{NS}(i)}{2}
\end{equation}
and,
\begin{equation}\label{eq:c2}
c_2(i)=\frac{2}{|E-V(i)|}\sin\left( \left[ \frac{|k(i)| \Delta x}{4} \right]^2 \right).
\end{equation}
Equations \ref{eq:ns1} and \ref{eq:ns2} are the non-standard form of the FDTD for the Schr\"odingier equation.  In the next section, we are going to include absorbing boundary conditions in order to simulate infinite systems.
\section{NS-FDTD with PML's}
To avoid reflections from the boundaries, we need some way to absorb the waves. The perfectly matched layers (PML's) where introduced in the electromagnetic version of the FDTD.\cite{SullivanBook} In a similar way, for the quantum mechanical version of the PML's, a stretching parameter needs to be introduced in the Schr\"odinger equation:
\begin{equation}
\frac{\partial \psi}{\partial t}=i\frac{\hbar}{2m}\left( \gamma \frac{\partial^2\psi}{\partial x^2} \right)-\frac{i}{\hbar}V\psi,
\end{equation}
where $\gamma$ is a complex quantity:
\begin{equation}
\gamma=\gamma_{re} + i \gamma_{im}.
\end{equation}
Details on $\gamma$ and related quantities, can be found elsewhere.\cite{Sullivan2012}
Again, separating $\psi$ into the real an imaginary parts, we obtain two coupled equations:
\begin{equation}
\frac{\partial \psi_{real}}{\partial t}=-\frac{\hbar}{2m}\left( \gamma_{real}\frac{\partial^2 \psi_{imag}}{\partial x^2}  +  \gamma_{imag}\frac{\partial^2 \psi_{real}}{\partial x^2}   \right) +\frac{V}{\hbar}\psi_{imag}
\end{equation}
and,
\begin{equation}
\frac{\partial \psi_{imag}}{\partial t}=\frac{\hbar}{2m}\left( \gamma_{real}\frac{\partial^2 \psi_{real}}{\partial x^2}  -  \gamma_{imag}\frac{\partial^2 \psi_{imag}}{\partial x^2}   \right) -\frac{V}{\hbar}\psi_{real}.
\end{equation}
Applying the finite difference scheme, these equations become:
\begin{equation}
\frac{\psi^n_{real} (i)-\psi^{n-1}_{real} (i)}{S(\omega,\Delta t)}=-\frac{\hbar}{2m}\frac{\gamma_{real} d_x^2\psi^{n-1/2}_{imag} (i)+ \gamma_{imag} d_x^2\psi^{n-2}_{real} (i)}{S(k,\Delta x)^2} + \frac{V(j)}{\hbar}\psi^{n-1/2}_{imag} (i)
\end{equation}
and,
\begin{equation}
\frac{\psi^n_{imag} (i)-\psi^{n-1}_{imag} (i)}{S(\omega,\Delta t)}=\frac{\hbar}{2m}\frac{\gamma_{real}d_x^2\psi^{n-1/2}_{real} (i) - \gamma_{imag} d_x^2\psi^{n-2}_{imag} (i)}{S(k,\Delta x)^2} - \frac{V(j)}{\hbar}\psi^{n-1/2}_{imag} (i).
\end{equation}
For the NS-FDTD algorithm, we can apply the same procedure in order to include the PML's.
After some rearrangement we obtain:
\begin{equation}
\psi^n_{real} (i)=\psi^{n-1}_{real} (i) -c_1(i) \left(  \gamma_{real} d_x^2\psi^{n-1/2}_{imag} (i) + \gamma_{imag} d_x^2\psi^{n-1/2}_{real} (i)  \right) + c_2(i)V(i)\psi^{n-1/2}_{imag} (i)
\end{equation}
and,
\begin{equation}
\psi^n_{imag} (i)=\psi^{n-1}_{imag} (i)+ c_1(i)\left(  \gamma_{real} d_x^2\psi^{n-1/2}_{real} (i)- \gamma_{imag} d_x^2\psi^{n-1/2}_{imag} (i)  \right) -c_2(i)V(i)\psi^{n-1/2}_{imag} (i)
\end{equation}
where $c_1$ and $c_2$ are given by eqs. \ref{eq:c1} and \ref{eq:c2}.
\section{Results}
\begin{figure}[h]
\begin{center}
\includegraphics[width=7.0 cm]{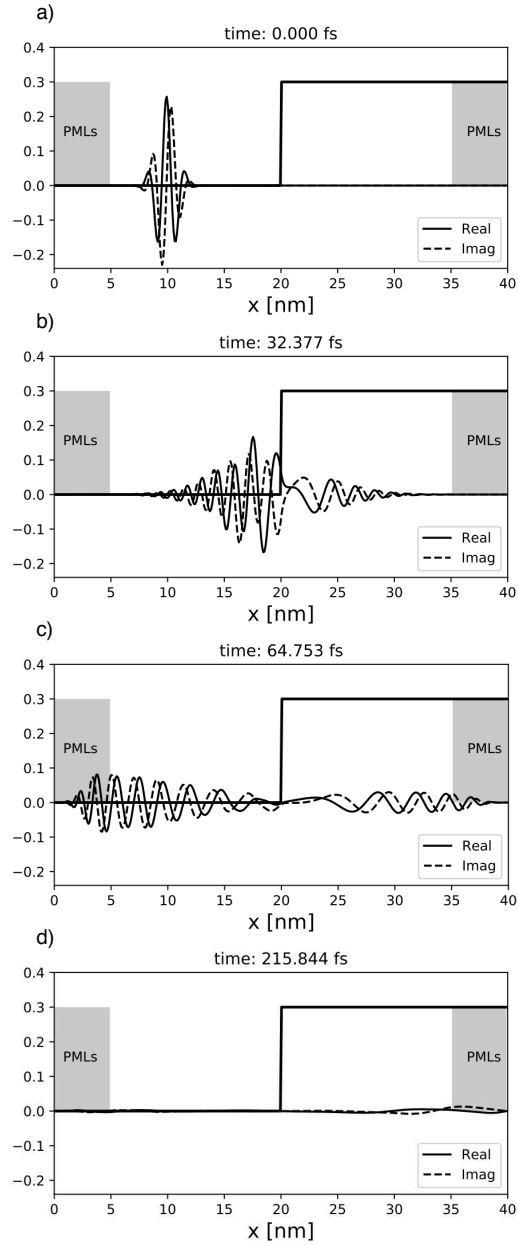}
\end{center}
\caption{Time snapshots of a wavefunction (real and imaginary parts) with an energy of $0.5$ $eV$ interacting with a step potential of $0.3$ $eV$. Transmission and reflection occurs as the wave reaches the potential and are absorbed in the PML's regions.}\label{fig:fig}
\end{figure}

In order to test our implementation, including the PML's version for the NS-FDTD, we present here the simulation of a step potential. In Fig. \ref{fig:fig} we show the time evolution of the real and imaginary parts of a wave function of $0.5$ $eV$. 
The wavelength is of $17$ \AA, the line of simulation has $40$ $nm$ of length, and the PML's have a length of $5$ $nm$ at the ends. Fig. \ref{fig:fig} a) shows the initial input for the wavefuction, being a Gaussian pulse of width of about the wavelength. At the time of about $32$ $fs$ in Fig. \ref{fig:fig} b) the pulse has reached the 0.3 $eV$ potential starting at the middle of the simulation line, and we can see the transmission and reflection of the wave. By the time of about $63$ $fs$ (Fig. \ref{fig:fig} c)), the reflected and transmitted waves are in the zone of the PML's and are absorbed as expected.  Finally, in Fig. \ref{fig:fig} d), at the time of about $216$ $fs$, most of the wave function has been almost completely absorbed.

\section{Conclusions}
The quantum mechanics version of the FDTD, has been treated in the methodology of the non-standard form proposed recently by Cole.\cite{Cole2013} The resulting iterative equations are different from the standard version, only in the form of the coefficients multiplying the kinetic and potential terms. The non-standard term depends on sine functions for $E>V$ and on a hyperbolic sine function for $E<V$ as the wavenumber is imaginary. However, the two forms are very similar. This means that any standard code can be easily converted to the NS-FDTD, by just using the coefficients given by eqs. \ref{eq:c1} and \ref{eq:c2}, and considering whether $E>V$ or $E<V$. In addition, we show that the PML's have also the same form, and we solve the time evolution of a one electron wave in the presence of a step potential including the absorbing boundary conditions. We hope that this new formulation of the FDTD applied to the Schr\"odinger equation can be used as a starting point to increase the precision of calculations for systems in two or three dimensions for simulating devices.

\bibliography{mybib}

\begin{thebibliography}{8}%
\makeatletter
\providecommand \@ifxundefined [1]{%
 \@ifx{#1\undefined}
}%
\providecommand \@ifnum [1]{%
 \ifnum #1\expandafter \@firstoftwo
 \else \expandafter \@secondoftwo
 \fi
}%
\providecommand \@ifx [1]{%
 \ifx #1\expandafter \@firstoftwo
 \else \expandafter \@secondoftwo
 \fi
}%
\providecommand \natexlab [1]{#1}%
\providecommand \enquote  [1]{``#1''}%
\providecommand \bibnamefont  [1]{#1}%
\providecommand \bibfnamefont [1]{#1}%
\providecommand \citenamefont [1]{#1}%
\providecommand \href@noop [0]{\@secondoftwo}%
\providecommand \href [0]{\begingroup \@sanitize@url \@href}%
\providecommand \@href[1]{\@@startlink{#1}\@@href}%
\providecommand \@@href[1]{\endgroup#1\@@endlink}%
\providecommand \@sanitize@url [0]{\catcode `\\12\catcode `\$12\catcode
  `\&12\catcode `\#12\catcode `\^12\catcode `\_12\catcode `\%12\relax}%
\providecommand \@@startlink[1]{}%
\providecommand \@@endlink[0]{}%
\providecommand \url  [0]{\begingroup\@sanitize@url \@url }%
\providecommand \@url [1]{\endgroup\@href {#1}{\urlprefix }}%
\providecommand \urlprefix  [0]{URL }%
\providecommand \Eprint [0]{\href }%
\providecommand \doibase [0]{http://dx.doi.org/}%
\providecommand \selectlanguage [0]{\@gobble}%
\providecommand \bibinfo  [0]{\@secondoftwo}%
\providecommand \bibfield  [0]{\@secondoftwo}%
\providecommand \translation [1]{[#1]}%
\providecommand \BibitemOpen [0]{}%
\providecommand \bibitemStop [0]{}%
\providecommand \bibitemNoStop [0]{.\EOS\space}%
\providecommand \EOS [0]{\spacefactor3000\relax}%
\providecommand \BibitemShut  [1]{\csname bibitem#1\endcsname}%
\let\auto@bib@innerbib\@empty
\bibitem [{\citenamefont {Yee}(1966)}]{Yee}%
  \BibitemOpen
  \bibfield  {author} {\bibinfo {author} {\bibfnamefont {K.}~\bibnamefont
  {Yee}},\ }\href@noop {} {\bibfield  {journal} {\bibinfo  {journal} {IEEE
  Transactions on Antennas and Propagation}\ }\textbf {\bibinfo {volume}
  {14}},\ \bibinfo {pages} {302} (\bibinfo {year} {1966})}\BibitemShut
  {NoStop}%
\bibitem [{\citenamefont {Sullivan}(2000)}]{SullivanBook}%
  \BibitemOpen
  \bibfield  {author} {\bibinfo {author} {\bibfnamefont {D.~M.}\ \bibnamefont
  {Sullivan}},\ }\href@noop {} {\emph {\bibinfo {title} {Electromagnetic
  Simulation Using the FDTD Method}}},\ edited by\ \bibinfo {editor}
  {\bibfnamefont {R.~J.}\ \bibnamefont {Herrick}}\ (\bibinfo  {publisher}
  {Wiley-IEEE Press},\ \bibinfo {address} {445 Hoes Lane, P. O. Box 1331,
  Piscataway, NJ 08855-1331},\ \bibinfo {year} {2000})\BibitemShut {NoStop}%
\bibitem [{\citenamefont {Sullivan}\ and\ \citenamefont
  {Citrin}(2003)}]{Sullivan2003}%
  \BibitemOpen
  \bibfield  {author} {\bibinfo {author} {\bibfnamefont {D.~M.}\ \bibnamefont
  {Sullivan}}\ and\ \bibinfo {author} {\bibfnamefont {D.~S.}\ \bibnamefont
  {Citrin}},\ }\href {\doibase 10.1063/1.1618916} {\bibfield  {journal}
  {\bibinfo  {journal} {Journal of Applied Physics}\ }\textbf {\bibinfo
  {volume} {94}},\ \bibinfo {pages} {6518} (\bibinfo {year}
  {2003})}\BibitemShut {NoStop}%
\bibitem [{\citenamefont {Soriano}\ \emph {et~al.}(2004)\citenamefont
  {Soriano}, \citenamefont {Navarro}, \citenamefont {Port{\'{i}}},\ and\
  \citenamefont {Such}}]{Soriano2004}%
  \BibitemOpen
  \bibfield  {author} {\bibinfo {author} {\bibfnamefont {A.}~\bibnamefont
  {Soriano}}, \bibinfo {author} {\bibfnamefont {E.~A.}\ \bibnamefont
  {Navarro}}, \bibinfo {author} {\bibfnamefont {J.~A.}\ \bibnamefont
  {Port{\'{i}}}}, \ and\ \bibinfo {author} {\bibfnamefont {V.}~\bibnamefont
  {Such}},\ }\href {\doibase 10.1063/1.1753661} {\bibfield  {journal} {\bibinfo
   {journal} {Journal of Applied Physics}\ }\textbf {\bibinfo {volume} {95}},\
  \bibinfo {pages} {8011} (\bibinfo {year} {2004})}\BibitemShut {NoStop}%
\bibitem [{\citenamefont {Sudiarta}\ and\ \citenamefont
  {Geldart}(2007)}]{Sudiarta2007}%
  \BibitemOpen
  \bibfield  {author} {\bibinfo {author} {\bibfnamefont {I.~W.}\ \bibnamefont
  {Sudiarta}}\ and\ \bibinfo {author} {\bibfnamefont {D.~J.~W.}\ \bibnamefont
  {Geldart}},\ }\href {\doibase 10.1088/1751-8113/40/8/013} {\bibfield
  {journal} {\bibinfo  {journal} {Journal of Physics A: Mathematical and
  Theoretical}\ }\textbf {\bibinfo {volume} {40}},\ \bibinfo {pages} {1885}
  (\bibinfo {year} {2007})}\BibitemShut {NoStop}%
\bibitem [{\citenamefont {Sullivan}\ and\ \citenamefont
  {Wilson}(2012)}]{Sullivan2012}%
  \BibitemOpen
  \bibfield  {author} {\bibinfo {author} {\bibfnamefont {D.~M.}\ \bibnamefont
  {Sullivan}}\ and\ \bibinfo {author} {\bibfnamefont {P.~M.}\ \bibnamefont
  {Wilson}},\ }\href {\doibase 10.1063/1.4754812} {\bibfield  {journal}
  {\bibinfo  {journal} {Journal of Applied Physics}\ }\textbf {\bibinfo
  {volume} {112}} (\bibinfo {year} {2012}),\ 10.1063/1.4754812}\BibitemShut
  {NoStop}%
\bibitem [{\citenamefont {Cole}\ \emph {et~al.}(2013)\citenamefont {Cole},
  \citenamefont {Okada},\ and\ \citenamefont {Banerjee}}]{Cole2013}%
  \BibitemOpen
  \bibfield  {author} {\bibinfo {author} {\bibfnamefont {J.~B.}\ \bibnamefont
  {Cole}}, \bibinfo {author} {\bibfnamefont {N.}~\bibnamefont {Okada}}, \ and\
  \bibinfo {author} {\bibfnamefont {S.}~\bibnamefont {Banerjee}},\ }\href
  {\doibase 10.1007/s12596-013-0134-0} {\bibfield  {journal} {\bibinfo
  {journal} {Journal of Optics (India)}\ }\textbf {\bibinfo {volume} {42}},\
  \bibinfo {pages} {316} (\bibinfo {year} {2013})}\BibitemShut {NoStop}%
\bibitem [{\citenamefont {{Visscher}}(1991)}]{Visscher1991}%
  \BibitemOpen
  \bibfield  {author} {\bibinfo {author} {\bibfnamefont {P.~B.}\ \bibnamefont
  {{Visscher}}},\ }\href {\doibase 10.1063/1.168415} {\bibfield  {journal}
  {\bibinfo  {journal} {Computers in Physics}\ }\textbf {\bibinfo {volume}
  {5}},\ \bibinfo {pages} {596} (\bibinfo {year} {1991})}\BibitemShut {NoStop}%
\end{thebibliography}%

\end{document}